\documentstyle[aps,multicol,epsfig]{revtex}

\begin{document}

\draft

\title{Stochastic control of quantum coherence}

\author{Stefano Mancini$^{1,2}$, David Vitali$^{1}$,
Paolo Tombesi$^{1}$, and Rodolfo Bonifacio$^{2}$}

\address{
$^{1}$INFM, Dipartimento di Fisica,
Universit\`a di Camerino,
I-62032 Camerino, Italy \\
$^{2}$ INFM, Dipartimento di Fisica,
Universit\`a di Milano,
Via Celoria 16, I-20133 Milano, Italy}

\date{\today}

\maketitle

\begin{abstract}
The application of a random modulation of a system parameter
usually increases decoherence effects. Here we show how, employing
an appropriate stochastic modulation, it is instead 
possible to preserve the quantum coherence of a system.
\end{abstract}


\begin{multicols}{2}

Controlling quantum coherence is one of the most fundamental issues in 
modern information processing \cite{QI}. 
The most popular solution in the
field of quantum information are quantum error correction codes
\cite{qecc} and error avoiding codes \cite{eac}, both based on 
encoding the state into 
carefully selected subspaces of a larger Hilbert space
involving ancillary systems. 
The main limitation of these strategies
for combatting decoherence is the 
large amount of extra space resources required \cite{steane}; in particular,
if fault tolerant error correction is also considered, the number of 
ancillary qubits enormously increases.

For this reason, other alternative 
approaches which do not require any ancillary resources have been 
pursued, and which may be divided into two main categories, 
according to the form of interaction 
with the system under study \cite{LLOYD00}. 
If the interaction is one way, so that the controller acts on the 
system without obtaining any information about its state,
then the controller is called ``open loop" \cite{VT99}.
By contrast, if the controller acts on the system on the basis of 
information that it obtains about the state of the system, then
it is called ``closed loop" \cite{LLOYD00,TV95}.
In standard open loop techniques, control of quantum dynamics is 
achieved 
through the application of suitably tailored, time-dependent and 
deterministic, driving 
forces. Here we want to extend open loop control strategies by 
considering the possibility of using {\em stochastic} parameter 
modulations for quantum control. 

The common 
wisdom is that whenever a system is subject to noise, the quality of 
the
dynamic control is degraded, and that quantum coherence in particular 
is rapidly destroyed \cite{GIU,BOTV}. Here we show that this is not 
generally true and that quantum decoherence can be significantly 
suppressed if an appropriately tailored stochastic modulation of a 
system parameter is used. This fact is illustrated in this letter by 
considering the simple case of 
the dynamics of a single radiation mode in a lossy cavity, 
but the results can be generalized.
In this open system, decoherence has a dissipative origin since it is 
due to photons' leakage out of the cavity, and the stochastic control 
strategy will be implemented by modulating the cavity length. 

Let us consider a single radiation mode with annihilation operator $a$
within a lossy cavity, whose characteristic frequency is
$\omega= n \pi c/L$, with $n$ an integer number,
$c$ the speed of light and $L$ the cavity length.
If photons' leakage occurs through 
a partially transmitting mirror, the decay rate 
will be given by $\gamma=c {\cal T}/2L$,
with ${\cal T}$ the mirror's transmittivity.

In the case of optical frequencies, thermal excitation from 
the environment of the continuum of modes outside the cavity
is negligible and the dynamics is well described by 
the master equation \cite{GAR}
\begin{equation}\label{ME0}
\dot\rho\equiv{\cal L}\rho
=-i\omega\left[a^{\dag}a,\rho\right]
+\gamma{\cal D}[a]\rho\,,
\end{equation}
where ${\cal D}[A]B\equiv ABA^{\dag}-\{A^{\dag}A,B\}/2$
describes photon decay 
into the vacuum. This decay is also responsible for the rapid decay of 
any eventual quantum coherence generated within the cavity 
\cite{WALLS}.

Let us now try to preserve the quantum coherence of the radiation mode
using an appropriate stochastic control strategy. 
In particular, we randomly modulate the cavity length, that is, 
$L\to L(t) = L_0/y(t)$ with $y(t)$ a positive stochastic process. 
This is equivalent to a simultaneous random modulation of both the 
frequency and the decay rate of the cavity. 
This stochastic modulation of the cavity length 
moreover yields a dynamics which is indistinguishable from that 
driven by the constant, unmodulated, Liouvillian superoperator 
${\cal L}_{0}=-i\omega_0\left[a^{\dag}a,\ldots\right]
+\gamma_0{\cal D}[a]\ldots\,$,
where the parameters $\omega_0$
and $\gamma_0$ are fixed,
in the presence of a {\em random} evolution time $t'$. In fact, 
one has
$$
\rho(t) = {\rm\bf T}\exp\left\{\int_{0}^{t}ds {\cal L}(s)\right\}\rho(0)
=\exp\left\{{\cal L}_{0}t'(t)\right\}\rho(0),
$$
where ${\rm\bf T}$ denotes time ordering, and
we have defined the stochastic evolution time 
$t'(t)=\int_{0}^{t}ds y(s)$.
This observation reminds the recently proposed 
model-independent approach
to decoherence in quantum mechanics \cite{BOTV,RB} 
in which the evolution 
time is regarded as a random variable.

To establish a 
connection between the randomized time evolution of 
Refs.~\cite{BOTV,RB} and the model of a cavity mode with a stochastically 
modulated cavity length 
we assume that the statistical properties 
of the cavity length modulation factor 
$y(t)$ are determined just by the Gamma probability 
distribution $P(t,t')$ for the random evolution time $t'$ of 
Ref.~\cite{BOTV,RB}.
To be more specific, we assume that at discrete times $t_{n}$ 
separated by a time interval $\Delta t$, independent
random variables $y(t_{n})$ are generated, e.g. by a computer, 
according to the distribution
\begin{equation}\label{eq:Pxi}
P(y)=\left(\frac{\Delta t}{\tau}\right)^{\Delta t/\tau}
\frac{y^{\Delta t/\tau-1}}
{\Gamma(\Delta t/\tau)}e^{-y\Delta t/\tau}\,.
\end{equation} 
The random number $y(t_{n})$ determines the ``instantaneous'' cavity length
$ L(t_n)=L_{0}/y(t_{n})$.
Equation (\ref{eq:Pxi}) 
is a Gamma probability distribution \cite{mandel}, 
with parameter $\Delta t/\tau$, where $\tau$ 
quantifies the strength of the fluctuations. In fact it is $\langle 
y(t_{n})\rangle =1$ and $\langle (y(t_{n})-1)^{2}\rangle =
\tau/\Delta t$.
Choosing the probability distribution (\ref{eq:Pxi}) 
means choosing a specific, uncommon way of modulating 
the cavity length. In fact it is easy to see \cite{BOTV,RB} that 
it implies a strongly non-Gaussian cavity length modulation,
which assumes Gaussian properties ($P(y)\simeq
\exp\left[-(y-1)^{2} \Delta t/2 
\tau\right]/\sqrt{2\pi\tau/\Delta t }$) 
in the limit $\Delta t/\tau \gg 1$ only. 

The above introduced discreteness in time, 
is dictated by the unavoidably finite rate of random number generation. 
Nevertheless, the time interval $\Delta t$ can be much 
smaller than the typical time scale upon which one observes
the system dynamics (this is essentially determined by 
$\gamma_{0}^{-1}$ times the inverse of the mean photon number). 
Hence, we can consider the independent
cavity length rescalings $y(t_n)$ as occurring continously in time, 
i.e. $\Delta t \to 0$. 
In the continuous approximation, $y(t)$ becomes a
{\em white, non-Gaussian} stochastic 
process, which can be rewritten as $y(t)=1+\xi(t)$, where
$\xi(t)$ is a zero-mean stochastic process, such that 
$\langle\xi(t)\xi(t')\rangle=\tau\delta(t-t')$.

As noted above, the evolution in the presence of
the cavity length modulation can be reinterpreted as the dynamics of 
a cavity with fixed length and with a random evolution time
$t'(t)$. However, since the sum (integral) of 
independent Gamma-distributed processes 
is again a Gamma-distributed process with the parameter given by
the sum (integral) of the single parameters \cite{mandel},
the probability distribution of the effective random time $t'(t)$ 
is just the Gamma distribution of Refs.~\cite{BOTV,RB},
\begin{equation}\label{PW}
P(t,t')=
\frac{e^{-t'/\tau}}{\tau}
\frac{[t'/\tau]^{t/\tau-1}}{\Gamma(t/\tau)}\,\,\,\,\,\,\,t'\geq 0,
\end{equation}
with parameter $t/\tau$. This fact simplifies the study of 
the dynamics of the dissipative radiation mode
in the presence of the stochastic modulation of the cavity length.
In fact any dynamical quantity can be obtained
by first evaluating it in the absence of modulation and then 
averaging it over the random time distribution (\ref{PW}).

A first interesting quantity is the time evolution of the 
intracavity mean photon number $\overline{\langle a^{\dag}(t)a(t) 
\rangle}=
\overline{\langle n(t) \rangle}$
(the time average is denoted by the overbar).
In the absence of any stochastic 
modulation one has $\overline{\langle n(t) \rangle}
= \langle n(t) \rangle=\langle n(0)\rangle
\exp(-\gamma_0 t)$, showing the energy decay due to 
photon leakage through the partially transmitting mirror.
In the presence of the cavity length modulation one has instead 
$$\overline{\langle n(t) \rangle}=\int\, dt'\, P(t,t') 
\langle n(t') \rangle=\langle n(0) \rangle 
e^{- \tilde\gamma_{e} t},$$
with the new effective energy decay rate $\tilde\gamma_{e} $
given by 
\begin{equation}
\tilde\gamma_{e} =\tau^{-1}
\log\left(1+\gamma_0\tau\right) \,. \label{gaen}
\end{equation}
It is always $\tilde\gamma_{e} \leq \gamma_{0}$ and therefore cavity 
length modulation always yields inhibition of dissipation. 
What is relevant is that the suppression of energy damping increases 
for increasing fluctuation strength parameter $\tau$ and that one has 
perfect inhibition, $\tilde\gamma_{e}=0$, in the limit 
$\gamma_{0}\tau \to \infty$, when the modulation becomes strongly 
non-Gaussian.

Another interesting quantity is the behavior of the
cavity field, which is essentially expressed by the average
$\overline{\langle a(t) \rangle}$. In the absence of any stochastic 
modulation one has $\overline{\langle a(t) \rangle}
= \langle a(t) \rangle=\langle a(0)\rangle
\exp(-i\omega_{0}t-\gamma_0 t/2)$. This quantity shows the effect of 
photon leakage on the phase of the intracavity field 
and it may provide some information on the possibility to control 
decoherence using cavity length modulation. In fact, since in this model
decoherence is just caused by photon leakage, we expect that any 
control exerted on the field decay rate will reflect itself into a 
control of quantum decoherence.
In the presence of the cavity length modulation one has
$\overline{\langle a(t) \rangle}=\langle a(0) \rangle 
e^{-i\tilde\omega t- \tilde\gamma t/2}$,
where the renormalized field decay rate $\tilde\gamma $
is 
\begin{equation}
\tilde\gamma = 
\tau^{-1}\log\left[\left(1+\gamma_0\tau/2\right)^2+\omega_0^2\tau^2
\right]\,, \label{garen}
\end{equation}
and the renormalized oscillation frequency is
$\tilde\omega  = \tau^{{-1}} \arctan\left[\omega_0\tau/(1+\gamma_0\tau/2)
\right] $.
The effective field decay rate $\tilde\gamma$ of Eq.~(\ref{garen}) 
is different from the effective energy decay rate $\tilde\gamma_{e}$
because it is sensitive not only to the decay rate modulation induced 
by the cavity length modulation (as $\tilde\gamma_{e}$), but also to 
the simultaneously induced frequency modulation, which may provide 
phase damping effects.
The behavior of the effective field decay rate (\ref{garen}), 
is plotted in Fig.~1
as a function of the modulation strength parameter $\gamma_{0}\tau$.
Curve (a) of Fig.~1 refers to Eq.~(\ref{garen}) and, at variance with 
what happens for $\tilde\gamma_{e}$ of Eq.~(\ref{gaen}),
shows an initial {\em increase} of the effective cavity decay rate
for increasing modulation amplitude $\tau$. This means that for not 
too large $\tau$, the modulation of the cavity length {\em increases} 
the cavity field decay rate.
This decay acceleration reaches a maximum around $\omega_{0}\tau 
\simeq 1$  
and then starts to decrease for increasing $\tau$.
What is interesting is that the ratio 
$\tilde\gamma /\gamma_{0}$ becomes less than one 
and tends to zero for larger $\tau$. 
This means that not only dissipation, but also field decay 
can be completely {\em inhibited} by the cavity length 
modulation at a sufficiently large $\tau$ parameter, when 
the stochastic modulation assumes strongly non-Gaussian properties.
The threshold value $\tau_{th}$ for decay inhibition,
$\tilde\gamma < \gamma_0$, depends on the cavity 
quality factor $Q=\omega_{0}/\gamma_{0}$, 
and it can be calculated iteratively, obtaining
\begin{equation}\label{eq:tauth}
\gamma_{0}\tau_{th}\approx
2\log Q+2\log\left[2\log Q\right] \,.
\end{equation}
Curve (b) of Fig.~1 instead refers to the cavity field decay rate
in the case of a Gaussian cavity length modulation, 
whose
expression essentially coincide with that obtained by extrapolating 
the expansion of Eq.~(\ref{garen}) 
at first order in $\tau$, that is, 
$\gamma_{Gaus}= \gamma_{0}+
\left(\omega_{0}^{2}-\gamma_{0}^{2}/4\right)\tau $.
The Gaussian modulation case always yields an accelerated decay rate 
(it is always $\omega_{0} \gg 
\gamma_{0}$ in optical cavities) showing that the 
statistical properties of the stochastic modulation (its non-Gaussian 
properties in particular) are a fundamental 
ingredient for achieving a significant decay inhibition. 

\begin{figure}[h]
\centerline{\epsfig{figure=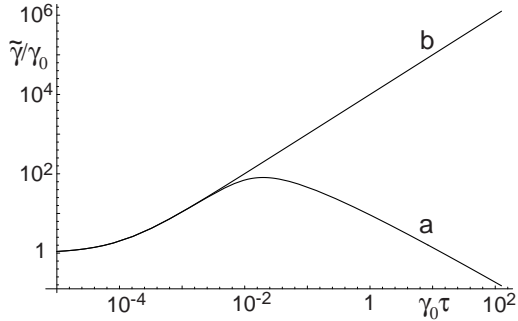,width=3.0in}}
\caption{\narrowtext Log-log plot of the ratio
$\tilde\gamma/\gamma_0$ as function of $\gamma_0\tau$.
Curve (a) refers to the Gamma stochastic modulation of
Eq.~(\protect\ref{eq:Pxi}), and curve (b) refers to a Gaussian 
stochastic cavity length modulation.
We have also set $Q=10^2$.}
\label{fig1}
\end{figure}

Let us now directly address the decoherence control issue.
We consider as initial state 
of the cavity field a linear superposition state. 
In order to control the coherence in the continuous variable case
we shall consider the
well known Schr\"odinger cat state, a superposition of two 
coherent states of the form 
$|\alpha\rangle+|-\alpha\rangle$  
and see what happens by employing the above stochastically modulated
dynamics. The same could eventually be done for a
superposition of Fock states. 

The time evolution of the Schr\"odinger 
cat state in the absence of any modulation is determined 
by the usual Liouvillian 
and it can be described in the following way \cite{WALLS}
$\rho(t)={\cal N}^{2}\{[|\alpha(t)\rangle\langle\alpha(t)|
+|-\alpha(t)\rangle\langle -\alpha(t)|]+\exp[-2|\alpha|^{2}(1-\eta(t))]
[|\alpha(t)\rangle\langle -\alpha(t)|
+|-\alpha(t)\rangle\langle \alpha(t)|]\}$\,,
where we have introduced 
$\alpha(t)=\alpha\exp[-(i\omega_{0}+\gamma_{0}/2)t]$
and $\eta(t)=e^{-\gamma_0 t}$.
A good characterization of the time development of the 
quantum coherence of the state of the cavity mode
is provided by the visibility with
respect to an observable \cite{WALLS}.
For the quadrature observable 
$X=(a+a^{\dag})/\sqrt{2}$, the 
quantum visibility is given by \cite{WALLS}
\begin{equation}\label{VXDEF}
{\cal V}=\frac{
\left|e^{-2|\alpha|^2[1-\eta(t)]}
\langle X|\alpha(t)\rangle
\langle -\alpha(t)|X\rangle\right|}
{\sqrt{\left[
\langle X|\alpha(t)\rangle
\langle\alpha(t)|X\rangle\right]
\left[
\langle X|-\alpha(t)\rangle
\langle-\alpha(t)|X\rangle
\right]
}}\,,
\end{equation}
where
$\langle X|\alpha\rangle=\left(\frac{1}{\pi}\right)^{1/4} 
\exp\left[-\frac{|\alpha|^2}{2}-\frac{X^2}{2}
-\frac{\alpha^2}{2}
+\sqrt{2}X\alpha\right]$.
In the absence of modulation, Eq.~(\ref{VXDEF})
leads to the simple result 
${\cal V}=\exp\left\{-2|\alpha|^2\left[1-\eta(t)\right]\right\}$.
In the presence of the stochastic modulation of the cavity length,
the corresponding visibility can be evaluated by performing an 
appropriate average of the dynamical quantities over the probability 
distribution $P(t,t')$. In particular, we have to 
consider the following replacements in Eq.~(\ref{VXDEF})
$e^{-2|\alpha|^2(1-\eta(t))}\langle X|\alpha(t)\rangle
\langle-\alpha(t)|X\rangle
\to
\overline{
e^{-2|\alpha|^2(1-\eta(t))}\langle X|\alpha(t)\rangle
\langle-\alpha(t)|X\rangle
}$,\,\,
$\langle X|\alpha(t)\rangle\langle\alpha(t)|X\rangle
\to\overline{
\langle X|\alpha(t)\rangle\langle\alpha(t)|X\rangle
}$,\,\,
$\langle X|-\alpha(t)\rangle\langle-\alpha(t)|X\rangle
\to\overline{
\langle X|-\alpha(t)\rangle\langle-\alpha(t)|X\rangle}$
to get the corresponding, averaged, $\overline{\cal V}$.
A cumbersome analytic expression can be obtained \cite{futur} and 
the corresponding behavior of $\overline{\cal V}$ as a function of 
time for different values of the modulation strength parameter $\tau$
is shown in Fig.~2. The relevant result is that the visibility, i.e., 
the quantum coherence properties of the system, {\em behaves in the same 
way as the field decay rate}. In particular we see either an acceleration, 
or, more importantly, a deceleration of decoherence according to 
the value of the parameter $\tau$.
The usual decay of the 
visibility in the absence of modulation ($\tau=0$) 
is shown with a dashed curve. 
When $\gamma_0\tau \neq 0$ we observe 
an acceleration of the decay of the visibility 
(lower curve) when the modulation strength $\tau$ is not too large
($\gamma_{0}\tau= 1.5$ in the figure) or a slowing down of the decay 
(upper curves) when $\tau$ becomes sufficiently large ($\gamma_{0}\tau= 
20,100$ in Fig.~2). The threshold value 
between the two behaviors coincides with that for decay inhibition 
$\tau_{th}$ of Eq.~(\ref{eq:tauth}).

\begin{figure}[h]
\centerline{\epsfig{figure=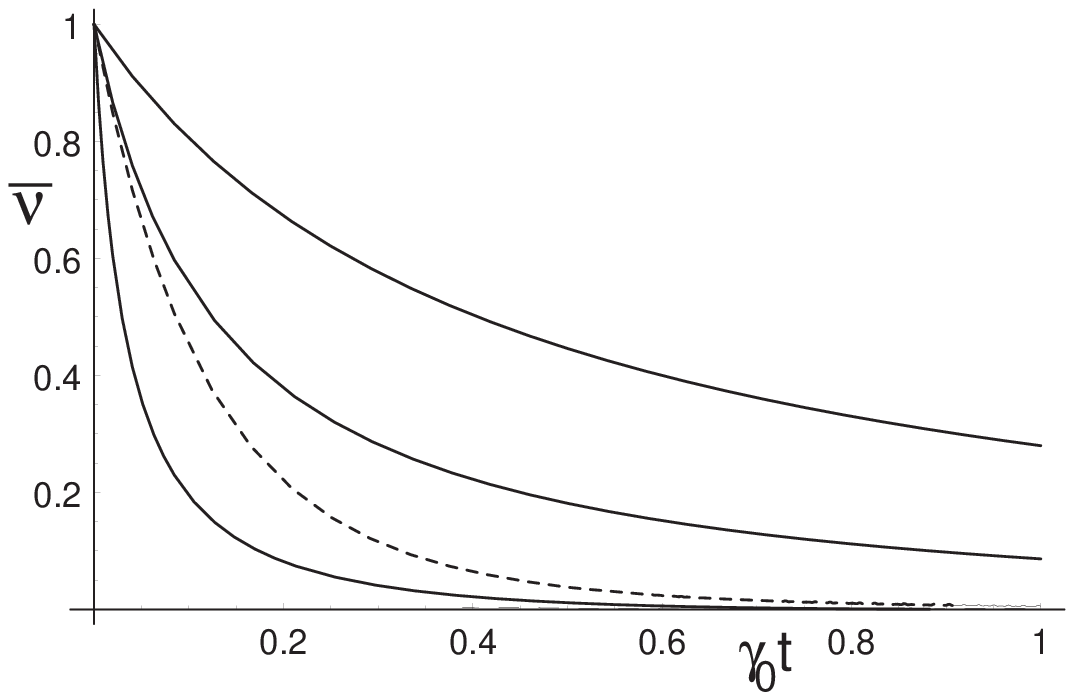,width=3.0in}}
\caption{\narrowtext
Time evolution of the visibility for different values 
of $\gamma_0\tau$. Curves from bottom  to top refer to 
$\gamma_0\tau=1.5$, $\gamma_0\tau=0$,
$\gamma_0\tau=20$, $\gamma_0\tau=10^2$ 
respectively. We have also used $\alpha=2i$.}
\label{fig2}
\end{figure}

The above results show that cavity dissipation, field decay, and 
decoherence can be 
inhibited if an appropriate random modulation of 
the cavity length is applied. We have assumed a Gamma distributed 
stochastic process (see Eq.~(\ref{eq:Pxi})) and we have seen that the 
inhibition becomes significant when the stochastic modulation becomes 
strongly non-Gaussian ($\gamma_{0}\tau $ large). Since the 
statistical nature of the modulation plays such a relevant role, it 
is important to establish the stability of the above decoherence control results 
with respect to small changes of the stochastic modulation.
As suggested above, 
the Gamma stochastic modulation of the cavity length
could be experimentally implemented using a cavity with a computer-controlled 
length, and a fast random number generator. Any eventual imperfect 
stochastic modulation of the cavity length can be described in terms 
of an {\em additional}, zero-mean, white, stochastic process $e(t)$, 
such that $\langle e(t)e(t')\rangle=\sigma\delta(t-t')$. The 
stochastic process $e(t)$ describes the ``error" in the modulation at 
time $t$, so that the actual fluctuating cavity length is 
$L(t)=L_{0}/(y(t)+e(t))$. We have checked that the above decay and 
decoherence inhibition results are not significantly changed as long 
as the strength $\sigma$ of the additional, undesired, modulation $e(t)$ is 
not too large. More precisely, it is possible to see with a 
perturbative treatment that the results are stable if
$
\sigma \gamma_{0}Q^{2} \ll 1 $.
This condition can be explicitely seen for example in the case of the 
renormalized field decay rate $\tilde\gamma$ and it is shown in 
Fig.~3 in the case of a Gaussian stochastic process $e(t)$.  

\begin{figure}[h]
\centerline{\epsfig{figure=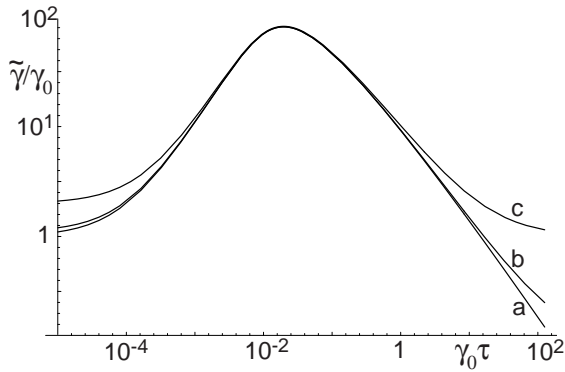,width=3.0in}}
\caption{\narrowtext Log-log plot of the ratio
$\tilde\gamma/\gamma_0$ as function of $\gamma_0\tau$
for different values of the ``error strength'' $\sigma$.
Curves are for $\sigma\gamma_{0}=0$ (a), 
$10^{-5}$ (b), $ 10^{-4}$ (c).
We have also set $Q=10^2$.}
\label{fig3}
\end{figure}

In conclusion, we have studied the possibility of a {\it stochastic
control} of (dissipative) decoherence by tailoring suitable 
random modulations of a system parameter. 
Against the widespread opinion that ``noise" is detrimental 
for quantum effects, 
we have shown that if the statistical properties of the modulation
are appropriately chosen, this stochastic control strategy
could be used in principle to control 
decoherence. Here we have considered the specific model of a single 
cavity mode with a randomly modulated cavity length. We 
have seen that, when the 
modulation is stochastic, with strongly non-Gaussian properties, 
decoherence and dissipation can be inhibited and that the
scheme can even tolerate some imperfection in the
modulation.
Although we have considered a specific model, 
our results can be generalized to dissipative system
in Ohmic environments where
the damping rate and the frequency have the same 
fluctuations \cite{GAR}. That allows us to recast
the above described treatment.
Finally, our approach shares some similarities
with the inhibition of atomic decay through random ac-Stark shift
discussed in Ref.~\cite{kurizki}. However our proposal 
is different since it strongly
depends on the statistical properties of the random modulation and
it is especially suited to the control of quantum decoherence.
Another analogy occurs with
the use of kicks to prevent the decay of a system \cite{VT99}.
In this latter case, dephasing introduced by kicks were deterministic 
processes well defined in time.
Instead, the present approach is merely probabilistic, so
it would be more manageable.

This work was partially supported by INFM under the PAIS {\it Entanglement 
and Decoherence} and by MIUR under the PRIN {\it Decoherence Control 
in Quantum Information Processing}.

\end{multicols}

\end{document}